\documentclass[twocolumn,aps,pra,showpacs,superscriptaddress,nofootinbib]{revtex4-1}
\usepackage{graphicx, color}% Include figure files
\usepackage{dcolumn}% Align table columns on decimal point
\usepackage{bm}% bold math
\usepackage{multirow}
\usepackage{amsmath}
\usepackage{epstopdf}
\usepackage{natbib}
\usepackage{soul}

\begin{document}

%\preprint{APS/123-QED}

\title{Contextuality in spin-orbit laser modes}% Force line breaks with \\
%\thanks{A footnote to the article title}%

\author{M. H. M. Passos$^{1,4}$, W. F. Balthazar$^2$, J. Acacio de Barros$^3$, C. E. R. Souza$^4$ A. Z. Khoury$^4$, J. A. O. Huguenin$^{1,4}$}

\affiliation{%
1- Instituto de Ci\^encias Exatas - Universidade Federal Fluminense - Volta Redonda - RJ - Brasil\\
2- Instituto Federal do Rio de Janeiro - Volta Redonda - RJ - Brasil\\
3- School of Humanities and Liberal Studies - San Francisco State University - San Francisco - CA - USA\\ 
4-Instituto de F\'{\i}sica - Universidade Federal Fluminense - Niter\'oi - RJ - Brasil\\
}%

%\collaboration{CLEO Collaboration}%\noaffiliation

\date{\today}% It is always \today, today,
             %  but any date may be explicitly specified
 
\begin{abstract}
We experimentally observed the violation of Kujula-Dzhafarov noncontextuality inequalities  by non-separable spin-orbit laser modes. Qubits are encoded on polarization and transverse modes of an intense laser beam. A spin-orbit non-separable mode was produced by means of a radial polarization converter (S-plate). The results show that contextuality  can be investigated by using spin-orbit modes in a simple way, corroborating recent works stating that such system can emulate single-photon experiments. Additionally, an improvement in the non-separable mode preparation allowed us to observe a greater violation of the Clauser-Horne-Shimony-Holt inequality for such system. The results are in very good agreement with the theoretical predictions of quantum mechanics.
% * <barros@sfsu.edu> 2018-06-15T08:37:38.825Z:
% 
% > Additionally, an improvement on the violation of CHSH for such system was also observed.
% This sentence seems odd, as we are pushing the idea that CHSH is not really a reasonable set of inequalities due to violations of marginal selectivity (no-signaling condition). For example, it is possible that S is greater for this new experiment, but that the violation of marginal selectivity is also greater, right? Could the sentence be phrased as an improvement in visibility or something similar? Just a thought. 
% 
% ^.
\end{abstract}

%\pacs{03.65.Yz, 03.67.Bg, 42.50.Ex}% PACS, the Physics and Astronomy
% Classification Scheme.
%\keywords{Suggested keywords}%Use showkeys class option if keyword
%display desired
\maketitle

%\tableofcontents

\section{\label{introd}Introduction}

In a famous paper, Kochen and Specker (KS) \cite{origcontx} challenged certain realistic models for quantum mechanics \cite{epr}. In it, they showed that a realistic theory that reproduced the outcomes of sets of observable properties had to be contextual, independently of the state of the quantum system. To do this, they produced a set of operators such that the observed properties of certain subsets of commuting operators (defining different experimental contexts) was inconsistent with the properties of the whole set, under the assumption that quantum properties do not change with context.  For a three-level system, Klyachko, Can, Binicioglu, and Shumovsky proposed a a state-dependent contextuality inequality  \cite{kbcs}. Experimental studies of quantum contextuality were performed by using different systems, such as  neutrons \cite{neutrons}, ion traps \cite{iot}, nuclear magnetic resonance \cite{rmn}, and photons \cite{photon1, photon2, photon3}.

Of course, entanglement and contextuality are related. Cabello discussed the link between Bell-type inequalities and noncontextual theories \cite{cabello1}. Entanglement and contextual quantum behavior of spin and orbital angular momentum was observed for twin photons \cite{padget1}.

In quantum computation, contextuality is an important ingredient in computation using states with only real amplitudes (rebits) \cite{rebits} and in computation based in measurements \cite{compmeasure}. For computation based on correlation, contextual correlations are an important ingredient for the deterministic evaluation of nonlinear Boolean functions \cite{compbcor}. The reliability of this computation was investigated and it was showed that for bipartite systems, CHSH correlation \cite{chsh} is a sufficient condition for reliable computation \cite{ernesto1}. Recent work presented the noncontextual wirings, a component that provide a class of contextuality-free operations \cite{wirings}, a very important features for quantum computation.   

Contextuality, according to Kochen and Specker  \cite{origcontx}, is the impossibility of consistently assigning values to physical properties associated to observables in a way that is independent of the experimental setup. As KS demonstrated, for any quantum system with dimension three or greater, it is possible to find a set of observables (in their example, 117 observables) corresponding to 0 or 1 eigenvalues (yes-no questions) that must be contextual. Cabello later on provided a simpler example of a contextual set of 18 observables in a Hilbert space of dimension four \cite{cabello2}. The idea of the KS proof is the following. Imagine we have an observable  $P_1$ that is simultaneously measured with  $P_2$, $P_3$, and $P_4$ in one experimental setup (context 1), but is also measured in another experiment with observables  $P_5$, $P_6$, and $P_7$  (context 2). The assumption that all such quantum observables $P_i$   have the same truth-value in all contexts lead to a logical contradiction. Therefore, it is not possible to consistently assign to a property $P_i$  in one context the same truth value in another context for all properties in the set. In other words, the algebra of quantum observables lead to contextuality. 

For systems with deterministic inputs and random outputs,  Kujala, Dzhafarov, and Larsson proposed a definition and measurement for contextuality \cite{kdl}.  A study on two binary systems, including entangled half-spin particles, based on Ref. \cite{kdl}, provided a set of four bound inequalities, called here Kajula-Dzhafarov (KD) inequalities, that are necessary and sufficient to verify contextuality \cite{kd2}. Violation of at least one of them constitute a sufficient condition to attest that the system is contextual. Here, we use KD inequalities to verify the contextuality in spin-orbit laser modes.

Linear optical systems have been largely used to explore quantum features by encoding qubits in degrees of freedom of the electromagnetic field. Connection between classical polarization optics and two-level quantum systems was established by the matrix formalism \cite{aiello1}. Polarization and transverse modes of a laser beam (spin-orbit modes) can be used to produce the so called non-separable modes \cite{topo} which present the same structure of two qubit entanglement. It was showed that such mode violated the Clauser-Horne-Shimony-Holt (CHSH) inequality \cite{bellnos}. By encoding qubits in amplitude and phase modulation in a laser beam a Bell-like inequality was also violated \cite{Ledesma}. Classical polarization states presented behavior of entangled states showing the strong correspondence of classical fields and quantum entangled states \cite{eberly1}.  Classical hypercorrelation in wave-optics was used to construct an analogy of quantum superdense coding \cite{kagawala}. Tripartite system was investigated by adding propagation path in spin-orbit modes degrees of freedom also violating the quantum-like the Mermin's inequality \cite{tripartitenos}. The entanglement generated by system-environment interaction was recently emulated by spin-orbit modes \cite{environ}. Several other tasks as quantum cryptography \cite{cript1, cript2}, quantum gates and algorithms \cite{lg1, lg2, lg3}, quantum games \cite{qg1, qg2}, and teleportation \cite{telep1, telep2, telep3} were investigated by using intense laser beam and linear optical circuits. The results of such studies are compatibles with then predicted by quantum mechanics and open a fundamental discussion on quantum-classical boundary \cite{eberly2}. As a pratical consequence of this established field of investigation, well succeed application of non-separable classical states on polarization metrology was reported \cite{aiello2}.

Contextuality was investigated also in a classical system by exploring a linear optical setup \cite{Ckbcs}. Polarization and path propagation of a laser beam were used to encode the analog of a tripartite system, called classical trit, to investigate the KCBS inequality. As a result, this system provided classical states that violated the KCBS inequality, showing analogy with contextual quantum systems.  

In this work we explored spin-orbi laser modes to investigate the existence of contextuality in bipartite systems by means of the Kujala--Dzahafarov (KD) inequality. Our results show that the maximally non-separable spin-orbit mode violates the KD inequality, a necessary condition to characterize such system as presenting contextuality. We also revisited the experiment of violation of CHSH inequality for spin-orbit modes \cite{bellnos} by improving mode preparation in order to achieve a higher violation of CHSH inequality.  The article is organized as follow. Section II presents the CHSH and Kujala-Dzhafarov inequalities for spin-orbit modes. The experimental setup and procedures are presented in Section III. Section IV contain results and  discussions. In section V we summarize our work.

\section{\label{qi} Spin-orbit laser modes and quantum-like inequalities}

In this section we explore the analogy between Bell's inequalities for quantum mechanics and their classical counterpart, known as spin-orbit inequalities \cite{bellnos,kagawala,tripartitenos}. Following Ref. \cite{Kaled}, the vector structure of electromagnetic field is common to quantum mechanics and classical optics, since the degrees of freedom of the electromagnetic field in a laser beam can be described by a vector product. In this way, we explore the vector nature of polarization and spatial modes (Hermitian-Gaussian mode of first order) to combine them and build the so called spin-orbit mode. 

The linear polarization vector can be described by the unit vectors $\hat{e}_V$  for vertical and  $\hat{e}_H$ for horizontal  polarization. The first-order of Hermitian-Gaussian modes, $HG_{01}(\vec{r})$ and $HG_{10}(\vec{r})$, can be described by direction of transverse mode, vertical and horizontal.  Therefore, $\{\hat{e}_H,\hat{e}_V\}$ and $\{HG_{10}(\vec{r}),HG_{01}(\vec{r})\}$ form a basis to describe the polarization and transverse mode states, respectively.

When the electromagnetic field can be written as a product between the transverse structure and a polarization vector ($\vec{E}(\vec{r})=\psi(\vec{r})\hat{e}\;$), we have the so called separable mode, which is analogue to the product state in quantum mechanics. 

\begin{figure}[!htp]
 \centering
 \includegraphics[scale=0.3,clip,trim=0mm 10mm 0mm 10mm]{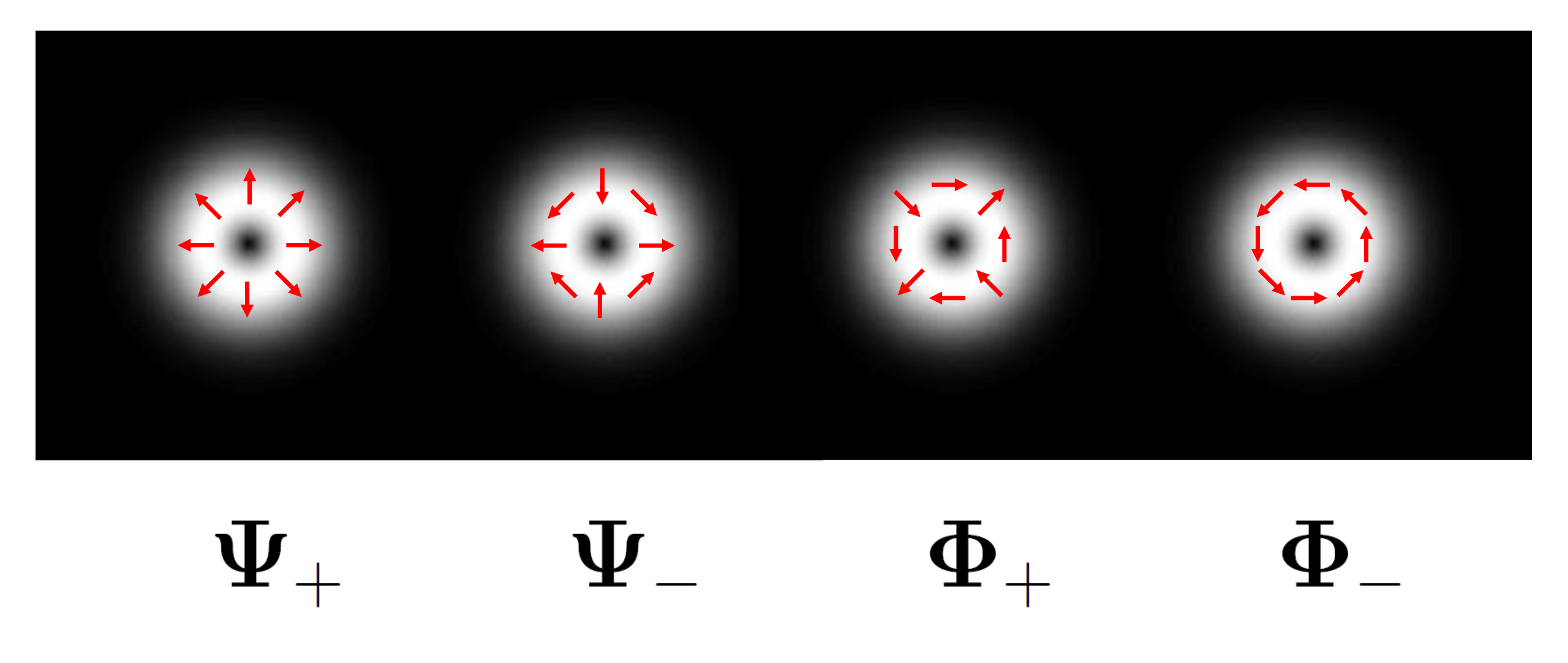}
 \caption{Transverse structure of maximally non-separable laser modes.The intensity has a donut shape. The arrows represent the polarization. $\Psi_\pm$ are the laser mode known as possessing radial polarization and $\Phi_\pm$ the azimuthal polarization.  }
 \label{vecbeams}
 \end{figure}

The most general vector of spin-orbit laser mode can be written as
\begin{eqnarray}
\vec{E}(\vec{r})= c_1  HG_{01}(\vec{r})\hat{e}_V  +  c_2HG_{01}&(\vec{r})&  \hat{e}_H  +  c_3HG_{10}(\vec{r})\hat{e}_V \nonumber \\ 
&+& c_4HG_{10}(\vec{r})\hat{e}_H,
\label{separable}
\end{eqnarray}
where  $c_i$, with $i=1,2,3$, and $4$, are the complex numbers satisfying the normalization condition. Here, normalization means that the sum of intensity of each term divided by the total intensity is equal to one. The values of complex coefficients $c_i$ can be choose to obtain a non-separable mode, that is non-factorable. We can write the spin-orbit laser modes as the radial ($\Psi_\pm$) and azimuthal ($\Phi_\pm$) polarization beams    
\begin{eqnarray}
\Psi_+(\vec{r})=\frac{1}{\sqrt{2}}\left[HG_{10}(\vec{r})\hat{e}_H    + HG_{01}(\vec{r})\hat{e}_V \right] ,\label{r1}\\
\Psi_-(\vec{r})=\frac{1}{\sqrt{2}}\left[HG_{10}(\vec{r})\hat{e}_H    - HG_{01}(\vec{r})\hat{e}_V\right]\label{r2}, \\
\Phi_+(\vec{r})=\frac{1}{\sqrt{2}}\left[HG_{10}(\vec{r})\hat{e}_V+HG_{01}(\vec{r})\hat{e}_H\right],\label{a12} \\
\Phi_-(\vec{r})=\frac{1}{\sqrt{2}}\left[HG_{10}(\vec{r})\hat{e}_V-HG_{01}(\vec{r})\hat{e}_H\right].
\label{az2}
\end{eqnarray}
As can be seen, this state cannot be written as the product of polarization and first order of Hermitian-Gaussian mode. Therefore, the maximally non-separable mode in Eqs.\ref{r1}-\ref{az2} present a vector structure that is similar to entangled states and can be associated with the Bell's basis. Indeed, $\{ \Psi_\pm, \Phi_\pm \}$ form an orthonormal mode basis. Figure \ref{vecbeams} presents the transverse structures of the maximally non-separable modes. The arrows distributed in the donut transverse structure illustrate radial and azimuthal polarization.

This analogy enable us to use the definition of the analogue to quantum Concurrence (C) to quantify the non separability \cite{topo}.  For a general mode, described by Eq.\ref{separable}, we can define
\begin{equation}
C = 2|c_1c_4-c_2c_3|.
\end{equation}
Note that for a maximally non-separable mode, Eqs.\ref{r1}-\ref{az2}, $C=1 $, which is equivalent to the Concurrence for the maximally entangled states of the Bell basis. Analogue to product states, $C=0$ for any separable mode. Consequently, due to such analogy, quantum-like inequalities for spin-orbit laser modes can be written and experimentally tested. 

\subsection{CHSH inequality}

Following  Ref.\cite{bellnos}, the same argument of Clauser-Horne-Shimony-Holt in quantum mechanics \cite{chsh} to demonstrate the spin-orbit inequality is used. For instance, we define the rotated basis of polarization and transverse modes, so that
\begin{align} 
\begin{split}
\hat{e}_{\alpha+}=(\cos \alpha)\hat{e}_V+(\sin \alpha)\hat{e}_H,  \\ 
\hat{e}_{\alpha-}=(\sin \alpha)\hat{e}_V+(\cos \alpha)\hat{e}_H, \\
HG_{+}(\vec{r})=(\cos \beta)HG_{01}+(\sin \beta)HG_{10}(\vec{r}),\\
HG_{-}(\vec{r})=(\sin \beta)HG_{01}-(cos \beta)HG_{10}(\vec{r}).
\end{split}
\end{align}
Eq.\ref{r1} can be rewritten in the rotated basis as
\begin{align}
\begin{split}
\Psi_{+}(\vec{r})=\frac{1}{\sqrt{2}} \cos(\beta-\alpha)\left[HG_{+}(\vec{r})\hat{e}_{\alpha+}+HG_{-}(\vec{r})\hat{e}_{\alpha-}\right]+\\
+\sin(\beta-\alpha)\left[HG_{-}(\vec{r})\hat{e}_{\alpha+}-HG_{+}(\vec{r})\hat{e}_{\alpha_-}\right].
\end{split}
\label{MNSbasis}
\end{align}
Defining  $I_{\pm\pm}$ as the normalized squared amplitude of each term of Eq.\ref{MNSbasis}, the normalized intensity of each component plays the role of probabilities in a quantum mechanics scenario. As Eq.\ref{MNSbasis} is normalized and the base vectors are unitary, we can shown that \cite{bellnos}
\begin{equation}
I_{++}(\alpha,\beta)+I_{+-}(\alpha,\beta)+I_{-+}(\alpha,\beta)+I_{--}(\alpha,\beta)=1.
\end{equation}
In addition, the quantity $M(\alpha,\beta)$ can be defined as \cite{bellnos}
\begin{equation}
\label{M(alpha,beta)}
M(\alpha,\beta) = I_{++}(\alpha,\beta) + I_{--}(\alpha,\beta) - I_{+-}(\alpha,\beta) - I_{-+}(\alpha,\beta),
\end{equation}    
and it is easy to show that 
\begin{equation}
\label{M(alpha,beta)}
M(\alpha,\beta) =cos\left[2\left(\beta-\alpha\right)\right],
\end{equation}    
which has the same result as the quantum mechanical correlations for a two-level system. The next step is to derive a Bell-type inequality for spin-orbit modes, given by
\begin{equation}
S = M(\alpha_1,\beta_1) + M(\alpha_1,\beta_2) - M(\alpha_2,\beta_1) + M(\alpha_2,\beta_2),
\label{S}
\end{equation} 
where $\alpha_k$ and $\beta_k$, with $k=1,2$, are two distinct angles. Again, as in quantum mechanics, for separable modes, $-2\leq S \leq2$, but for non-separable modes the Bell-type inequalities can be violated. For a set of angles, $\alpha_1=\pi/8$, $\alpha_2=3\pi/8$, $\beta_1=0$ and $\beta_2=\pi/4$, we obtain the maximum violation predicted by inequality, $S=2\sqrt{2}$. This inequality can be maximally violated by all modes presented in Eqs.\ref{r1}-\ref{az2}.

\subsection{\label{KDine} Kujala-Dzhafarov inequality }

Satisfaction of the CHSH inequalities is a necessary and sufficient condition for the existence of a joint probability distribution for all experimental outcomes. However, as it is well known, they rely on a major assumption, namely that the non-signaling condition is valid. Yet, due to many experimental factors, photon-correlation experiments often exhibit some form of violation of the non-signaling condition \cite{aspectdr}, with this condition being satisfied recently only in loophole free Bell-type experiments \cite{loop1, loop2} . 

Let us consider the standard Bell-EPR setup, where we have two settings for Alice and Bob's detectors, and four random variables representing the outcomes of measurements, often labeled $A_1$, $A_2$, $B_1$, and $B_2$, for Alice and Bob, respectively. It is a consequence of probability theory that $\langle A_1\rangle $ and $\langle A_2\rangle $  are independent of whether they were measured in conjunction with $B_1$ or $B_2$. However, if the non-signaling condition does not hold, this means that the intensity at the detector is not independent on the choices of the other detectors, and there is no possibility of having random variables (and therefore an underlying joint probability distribution) for $A_1$, $A_2$, $B_1 $, and $B_2$. Consequently, the use of the CHSH inequalities to determine contextuality (through the non-existence of a joint probability distribution) is inadequate, because a joint probability already does not exist. Therefore, to examine contextuality, we need a different set of inequalities from CHSH. 

Dzhafarov and Kujala \cite{kd2} provided a consistent framework within classical probability theory to describe general systems that are contextual. Here we will focus on the Bell-EPR setup to present their approach and inequalities. Since, because of violations of non-signaling, the statistical distribution of property $A_1$ when measured with $B_1$ is different from when it is measured with $B_2$, we will label its corresponding random variable as $A_{11}$ for the latter and $A_{12}$ for the former, with the first index indicating that we are measuring $A_1$ and the second index indicating the context $B_1$ or $B_2$ respectively. With such indexing, we have eight random variables, instead of the original four, namely $A_{11}$, $A_{12}$, $A_{21}$, $A_{22}$, $B_{11}$, $B_{12}$, $B_{21}$, and  $B_{22}$. The non-signaling condition would be represented, in this notation, by the requirement that $p(A_{ij}=A_{ij'})=1$ and $p(B_{ij}=B_{ij'})=1$ for $i,j,j'=1,2$ and $j\neq j'$. 

Of course, because they are measured together, the random variables $A_{ij}$ and $B_{jk}$ are jointly distributed. Furthermore, under no additional assumptions, $A_{ij}$  and $B_{kl}$ are not jointly distributed, but it is always possible to create a new set of jointly distributed random variables $A'_{ij}$  and $B'_{kl}$ such that all the stochastic properties of  $A_{ij}$  and $B_{kl}$ are reproduced. Such a new set of random variables is called in probability theory a coupling, since it connects variables from two different contexts as a single joint probability distribution. 

Because such couplings always exist, in fact an infinite number of them, providing a coupling is not sufficient to tell us whether the physical system is contextual. To address the question of contextuality, we need to look at all possible couplings under some additional condition. Let us, for example, consider $A_{11}$ and $A_{12}$, which are two random variables measured in different contexts and therefore not jointly distributed. When constructing a coupling, we can make whatever assumptions we want about the connection between those two variables. But if we want to think of them as the same property being measured in different contexts, we should try to construct a coupling such that the probability that they are the same is maximal-- being one is not a possibility, since their distributions are different due to artifacts that may cause a violation of the non-signaling violations. We can also take the same approach when trying to find a coupling that involves all variables. This constraint, i.e. the requirement the coupling is compatible with the maximal probability that a random variable representing a property in one context is equal to the same property in another context,  is called the multimaximal coupling \cite{kd1}. Now, even though a coupling always exist that reproduces the marginals, a multimaximal coupling does not always exist, which means that a joint probability distribution with a multimaximal coupling does not exist. When this is the case, the system is said to be contextual\footnote{Kaszlikowski and Kurzynski \cite{pawel}  point out that violation of the non-signaling condition is already a form of context dependency, which they call \textit{strong contextuality}, whereas the contextuality resulting from a lack of a joint they call \textit{hidden contextuality}, or simply \textit{contextuality}.}

For the standard Bell-EPR case, with the notation shown above, we are now in a position to write a set of inequalities that, if satisfied, guarantee that the system is non-contextual. Given the random variables $A_{ij},B_{kl}$, this system  is non-contextual if and only if it satisfies the following inequalities,  
% * <barros@sfsu.edu> 2018-07-10T00:42:06.856Z:
% 
% > Given the random variables $A_{ij},B_{kl}$, this system  is non-contextual if and only if it satisfies the following inequalities,  
% Would it make more sense to define S_KD etc as |<A11B11>+...-<A22B22>| - 2\Delta_0? This way we could have a direct comparison with the CHSH inequality, as it would be bound by 2.  This is different from what they write, but I think it would be clearer. 
% 
% ^.
\begin{eqnarray}
S_{KD1}\; & = & \;\label{eqcontex1}
\left|  \left< A_{11} B_{11} \right> + \left< A_{12} B_{12} \right> + \left< A_{21} B_{21} \right> - \left< A_{22} B_{22} \right> \right| \nonumber \\ 
&\leq & 2(1 + \Delta_0),
\end{eqnarray}
\begin{eqnarray}
\label{eqcontex2}
S_{KD2}\; & = & \; \left|  \left< A_{11} B_{11} \right> + \left< A_{12} B_{12} \right> - \left< A_{21} B_{21} \right> + \left< A_{22} B_{22} \right> \right|\nonumber \\
& \leq & 2(1 + \Delta_0),
\end{eqnarray}
\begin{eqnarray}
\label{eqcontex3}
S_{KD3}\; & = & \; \left|  \left< A_{11} B_{11} \right> - \left< A_{12} B_{12} \right> + \left< A_{21} B_{21} \right> + \left< A_{22} B_{22} \right> \right|\nonumber\\
& \leq & 2(1 + \Delta_0),
\end{eqnarray}
and
\begin{eqnarray}
\label{eqcontex4}
S_{KD4}\; & = & \; \left| - \left< A_{11} B_{11} \right> + \left< A_{12} B_{12} \right> + \left< A_{21} B_{21} \right> + \left< A_{22} B_{22} \right> \right| \nonumber \\
& \leq & 2(1 + \Delta_0),
\end{eqnarray}
where $\Delta_0$ is given by
\begin{eqnarray}
\label{d0}
\Delta_0  = & \frac{1}{2} & ( |\left< A_{11} \right> - \left< A_{12} \right>| + |\left< A_{21} \right>  -  \left< A_{22} \right>| \nonumber \\ 
&+&|\left< B_{11} \right> - \left< B_{21} \right>| + |\left< B_{12} \right> - \left< B_{22} \right>| ). 
\end{eqnarray}

Notice that these inequalities are the same as CHSH if we set $A_{ij}=A_{ik}$, $B_{ij}=B_{ik}$, and they are a generalization of CHSH for a system that violates the no-signaling condition. We can see that the terms on the right hand side of $\Delta_0$ are, intuitively, a measure of how different the expectations of each random variable are from one context to the other, and $\Delta_0$ is the cumulative sum of all such possible terms; the more no-signaling violation, the greater the value of $\Delta_0$. $\Delta_0$ has a very important role, as it changes the bounds of the inequalities that define the existence of a joint probability distribution under the assumption of a multimaximal coupling. 
%\begin{equation}
%\label{Sc}
%S_c = \left|  \left< A_{11} B_{11} \right> + \left< A_{12} B_{12} \right> - \left< A_{21} B_{21} \right> + \left< A_{22} B_{22} \right> \right| = 2.503
%\end{equation}

Considering spin-orbit modes system, we can write the mean values of KD inequalities as combination of normalized intensity $I_{\pm \pm} (\alpha_i, \beta_j)$ in the same way presented for CHSH inequality. For the calculation of $\Delta_0$, the average $\left< A_{i,j} \right>$ and $\left< B_{i,j} \right>$ are given by
\begin{eqnarray}
\label{Aij}
\left< A_{i,j} \right> &=& I_{++}(\alpha_i,\beta_j)\; - \; I_{--}(\alpha_i,\beta_j)\; + \;  I_{+-}(\alpha_i,\beta_j)\;\nonumber \\
&-&I_{-+}(\alpha_i,\beta_j)
\end{eqnarray}
and
\begin{eqnarray}
\label{Bij}
\left< B_{i,j} \right> &=& I_{++}(\alpha_i,\beta_j)\; - \;I_{--}(\alpha_i,\beta_j)\; - \;I_{+-}(\alpha_i,\beta_j)\;\nonumber \\
&+& I_{-+}(\alpha_i,\beta_j),
\end{eqnarray}
respectively. The average of the joint measurements are obtained by   
\begin{eqnarray}
\label{AijBij}
\left< A_{i,j} B _{i,j} \right> &=& I_{++}(\alpha_i,\beta_j)\; +\; I_{--}(\alpha_i,\beta_j)\; -\; I_{+-}(\alpha_i,\beta_j)\; \nonumber \\
&-& I_{-+}(\alpha_i,\beta_j),
\end{eqnarray}
where $i = 1, 2$, and $j = 1 , 2$ in Eqs.\ref{Aij}, \ref{Bij}, and \ref{AijBij}. It is worth to mention that KD inequalities construction use the same basis that ones used to construct CHSH inequality. In addition, for spin-orbit modes we can verify KD inequalities with the same normalized intensities used for CHSH experiment. Then, the same apparatus can be used to study both CHSH and KD inequalities. The experimental investigation of both of them is presented in the next section. 

\section{\label{exp} Experimental violation of quantum-like inequalities with spin-orbit modes}

The experimental study of the inequalities discussed in previous section by using spin-orbit modes were performed following the proposal presented in Ref.\cite{bellnos}, by improving the mode generation. The two qubits system was encoded in polarization and transverse mode of a laser beam. The first qubit was encoded in polarization, where horizontal (H) and vertical (V) polarizations represent the basis states $\left| 0 \right\rangle_P$ and $\left| 1 \right\rangle_P$, respectively. The second qubit was encoded in the first order of Hermite-Gauss modes. Thereby we have the HG$_{10}$, named h, and the HG$_{01}$, named v, representing the basis states $\left| 0 \right\rangle_M$ and $\left| 1 \right\rangle_M$, respectively. Table \ref{codif} summarizes the codification. It is worth to mention that we are using brackets notation to represent the laser beams modes due the analogy between laser modes and quantum states \cite{bellnos, aiello1, eberly1, kagawala, eberly2, Kaled}.
% * <barros@sfsu.edu> 2018-07-10T00:09:46.014Z:
% 
% > Thereby we have the HG$_{10}$, named h, and the HG$_{01}$, named v, representing the basis states $\left| 0 \right\rangle_M$ and $\left| 1 \right\rangle_M$, respectively. Table \ref{codif} summarizes the codification.\\
% 
% The definition above has what seems like an inconsistent notation with the rest of the text. In the text below nowhere |00>, |01> etc show up, but instead |Hh>, |Hv> etc are used. Should't this be fixed?
% 
% ^.
\begin{table}
\centering
\caption{Codification of qubits on the polarization and transverse mode HG of a laser beam.}
\label{codif}
\begin{tabular}{|c|c|}
\hline
\textbf{Optical modes basis}          & \textbf{Logical computational basis} \\ \hline
$\hat{e}_H$ HG$_{10}$ = $\left| Hh \right>$ & $\left| 00 \right>$                \\ \hline
$\hat{e}_H$ HG$_{01}$ = $\left| Hv \right>$ & $\left| 01 \right>$                \\ \hline
$\hat{e}_V$ HG$_{10}$ = $\left| Vh \right>$ & $\left| 10 \right>$                \\ \hline
$\hat{e}_V$ HG$_{01}$ = $\left| Vv \right>$ & $\left| 11 \right>$                \\ \hline
\end{tabular}
\end{table}

\begin{figure}[!htp]
 \centering
 \includegraphics[scale=0.4,clip,trim=0mm 45mm 0mm 30mm]{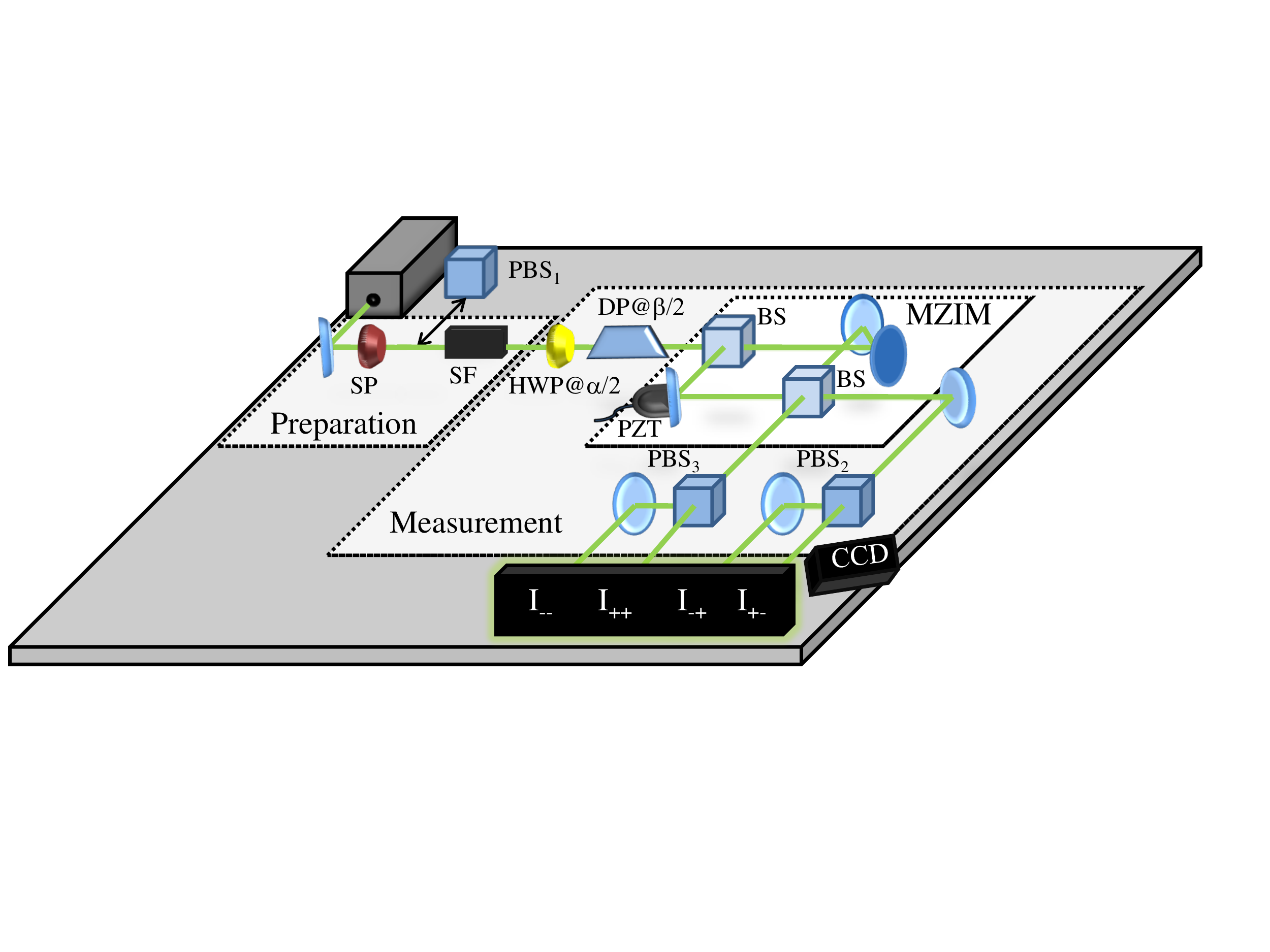}
 \caption{Experimental setup for the CHSH and Kujala-Dzhafarov Inequality. SP stand for S-wave plate, HWP for half wave plates, DP for Dove prism, BS for beam splitter, and PBS for polarizing beam splitter.}
 \label{expe.}
 \end{figure}

The experimental setup is illustrated in Figure \ref{expe.}. The state preparation starts with a diode pumped solid state (DPSS) laser beam (532 mm, 1.5 mW, and vertically polarized) illuminating a S-wave plate (SP). This device can be adjusted to produce directly the maximally non-separable mode 
\begin{equation}
\left| \Psi_- \right> = \frac{1}{\sqrt{2}} \left( \left| Hh \right> - \left| Vv \right>  \right),
\end{equation}
an analog of a two qubit entangled state, namely a maximally nonseparable state \cite{bellnos}. By introducing the PBS$_1$ in the laser path the component $\left| Hh \right>$ is transmitted, and we have the analog of a product state, namely a separable mode \cite{bellnos}. The preparation stage finishes with a spatial filter (SF) used to performs a mode clean in order to improve the modes fidelity. The mode preparation is the main experimental improvement introduced here in comparison to Ref.\cite{bellnos}.     
% * <barros@sfsu.edu> 2018-07-10T00:16:19.625Z:
% 
% >  (SF) used to performs a mode clean in order to improve the modes fidelity. 
% 
% SF is not in the figure. Should it be? 
% 
% ^.

The Bell-like measurement was performed by using a half-wave plate (HWP@ $\alpha_i$/2) and a Dove Prism (DP@$\beta_j$/2) to chose the measurement basis ($\hat{e}_{\alpha\pm}$, HG$_{\beta\pm}$), that are the orientation of the Bell basis measurement. The half-wave plate change the polarization states and the Dove Prims allowed us manipulate the transverse spatial mode. Therefore, after the combined action of this two devices,  the spin-orbit mode can be written as
\begin{eqnarray}
\label{mns}
\left| \Psi \right> = &\frac{1}{\sqrt{2}}& ( \cos(\beta - \alpha) \left| Hh \right>  + \cos(\beta - \alpha) \left| Vv \right>\nonumber \\ &+& \sin(\beta - \alpha) \left| Vh \right> - 
\sin(\beta - \alpha) \left| Hv \right>).
\end{eqnarray}
The projective measurements start with a Mach-Zehnder Interferometer with an additional mirror (MZIM) \cite{mzim}.
The MZIM performs a parity selection on the spin-orbit mode by adjusting the phase difference between interferometer arms, here implemented by a piezoelectric ceramic (PZT). Thereby, we have the even modes ($\left| Hh \right>$ and $\left| Vv \right>$) and the odd modes ($\left| Hv \right>$ and $\left| Vh \right>$) leaving the MZIM in different outputs. After the MZIM, outputs polarizing beam splitters (PBS$_2$ and PBS$_3$) project the four components in a bulkhead and the four intensities are registered simultaneously with a Charge Coupled Device (CCD) camera. Odd modes arrive in PBS$_2$, the component with the horizontal polarization $\left| Hv \right>$ (associated with I$_{+-}$ intensity) is transmitted and the component with the vertical polarization $\left| Vh \right>$ (I$_{-+}$) is reflected. Even modes arrive in PBS$_3$ and the  component with the horizontal polarization $\left| Hh \right>$ (I$_{++}$) is transmitted while the component with the vertical polarization $\left| Vv \right>$ (I$_{--}$) is reflected. It is worth to mention that to calculate M($\alpha_i$, $\beta_j$), $\left< A_{i,j} \right>$, $\left< B_{i,j} \right>$, and $\left< A_{i,j} B _{i,j} \right>$ we take the normalized intensities  I$_{\pm \pm}$, obtained by integrating the intensity distribution of each outputs images  divided by their sum.  In order to verify the violation of CHSH and Kujala--Dzhafarov inequalities we take the angles sets that corresponds to the maximal violation of CHSH inequality $\alpha_1 = \pi/8$, $\alpha_2 = 3\pi/8$, $\beta_1 = 0$, and $\beta_2 = \pi/4$.

% 	\begin{figure}[!htp]
% 		\centering
% 		%\includegraphics[scale=0.45,trim=3.4cm 0cm 0cm 0cm, clip=true]{figure1.eps}
% 		\includegraphics[scale=30,trim=0.05cm 0cm 0cm 0cm, clip=true]{figure1.eps}
% 		\caption{Experimental setup.}
% 		\label{Circuit}
% 	\end{figure} 

% \begin{figure}[!htp]
% \centering
% \includegraphics[scale=1.0]{figure2.eps}
% \caption{Double input/output MZIM.}
% \label{domzim}
% \end{figure}

\section{\label{results} Results and discussions}

\begin{figure}[htp!]
 \centering
 \includegraphics[scale=0.53,clip,trim=37mm 15mm 0mm 10mm]{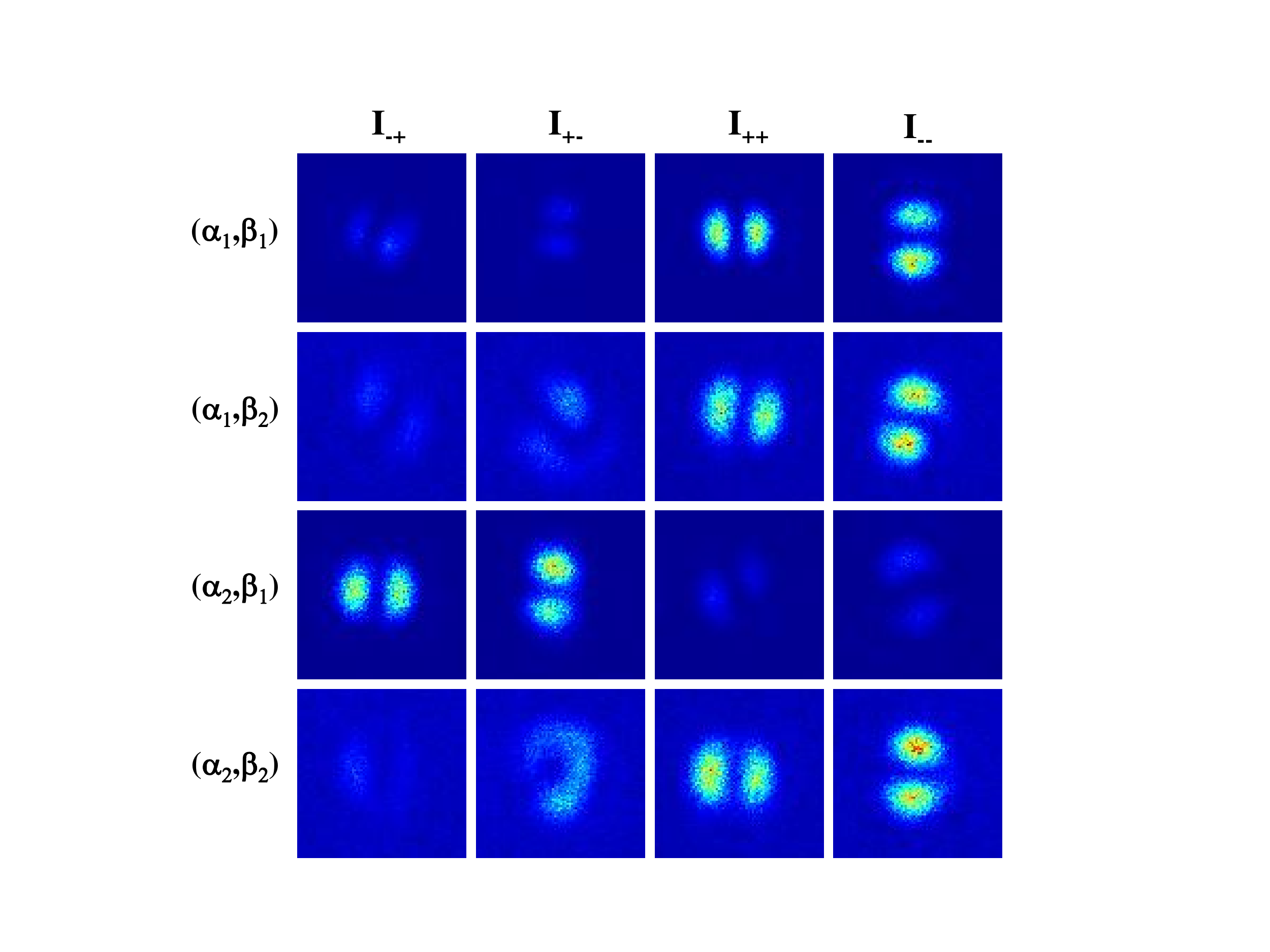}
 \caption{Resulting images for maximally non-separable mode.}
 \label{nonsepmodes}
\end{figure}

\begin{figure}[htp!]
 \includegraphics[scale=0.53,clip,trim=35mm 15mm 0mm 10mm]{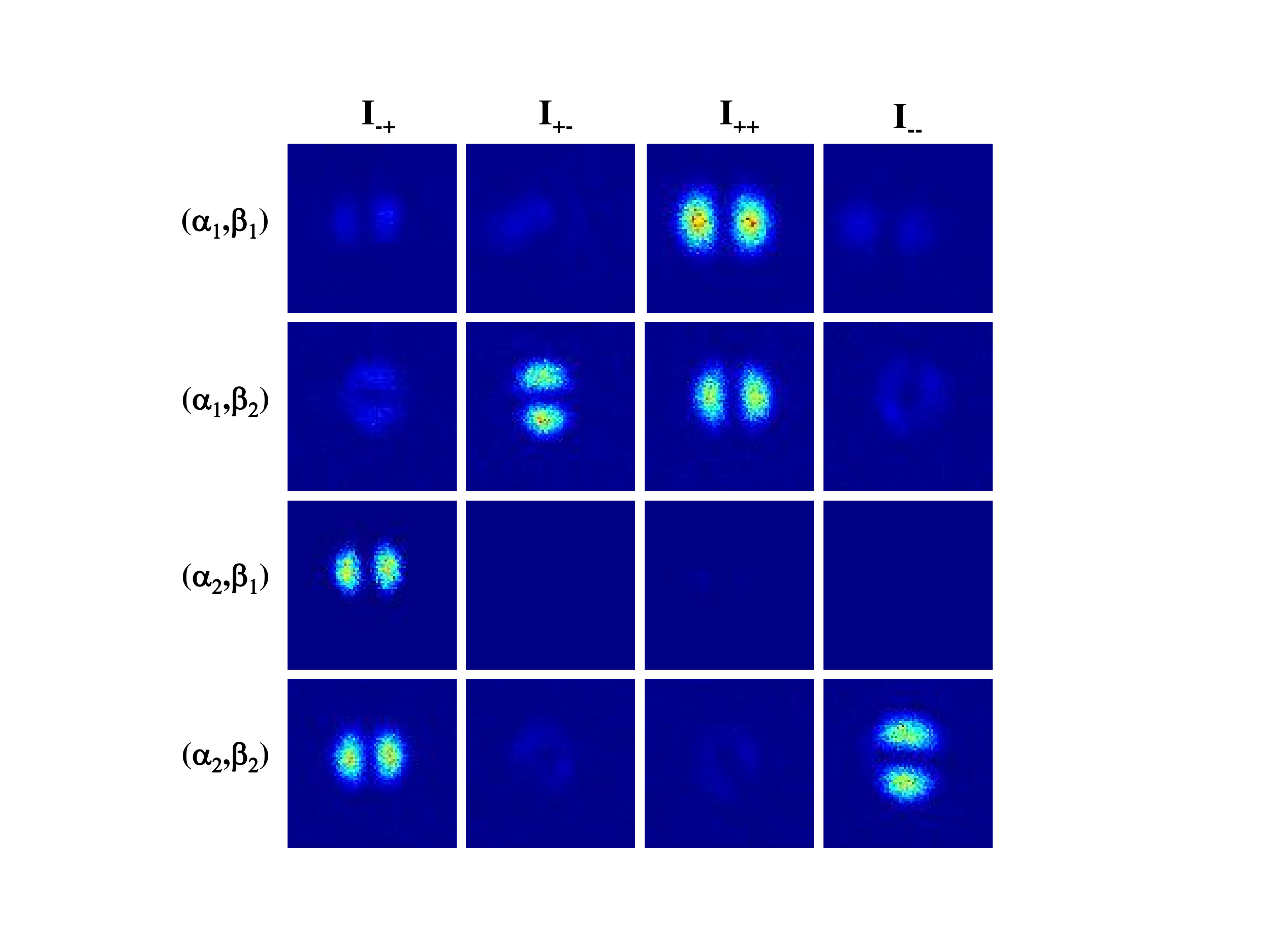}
 \caption{Resulting images for separable modes.}
 \label{sepmodes} 
\end{figure}

We performed the  experiment by preparing two initial state: a maximally nonseparable mode and a separable mode, as described in Section III.  By using the sets ($\alpha_i$, $\beta_j$) for maximal violation of CHSH inequality, we register the four intensities for each combination. Fig. \ref{nonsepmodes} shows the resulting intensities for the maximally nonseparable mode. The outputs of the measurement setup are identified by the respective intensity labels $I_{++}$, $I_{+-}$, $I_{-+}$, and $I_{--}$. Each row is equivalent to the simultaneously captured images for a combination of ($\alpha_i$, $\beta_j$). We can observe a good visibility. We found 90$\%$ for the combinations with $\beta_1$. For the combinations containing $\beta_2$, where the DP is rotated, the visibility is worse (85$\%$). The rotated DP slightly affects the polarization in a such way to transform a $\left|Hh\right>$ into $\left|Vh\right>$ and $\left|Vv\right>$ into $\left|Hv\right>$ mode components, making it hard to align the MZIM and consequently arrive at a better visibility.    
Fig.\ref{sepmodes} shows the resulting intensities for the separable mode $\left|Hh\right>$, obtained by including the PBS$_1$ in the laser path. The notations for intensities and angles are the same. The visibility of the MZIM for the separable mode is superior to the maximally non-separable case once the mode has only the $\left|Hh\right>$ component. Combinations with $\beta_2$ have also a worse visibility (around 88$\%$) compared to the ones containing $\beta_1$ (95$\%$).  The normalized image intensities were used to calculate the inequalities  described in the previous section.

\subsection{CHSH inequality with spin-orbit modes}

Table \ref{value_M} presents  both theoretical and experimental results for the calculation of $M(\alpha_i, \beta_j)$ and $S$. Theoretical values were obtained from the calculations of $M$ by using maximally entangled and product quantum states. Experimental values were obtained from the calculations of Eqs. \ref{M(alpha,beta)} and \ref{S}. For the maximally non-separable mode we observed a violation of $S=2.503$, an important improvement compared with the result of Ref.\cite{bellnos}. By comparing theoretical and experimental $M(\alpha_i, \beta_j)$ values we observe a more accentuated difference for the combinations containing $\beta_2$ due the limited visibility of the MZIM.    
\begin{table}[htp!]
\caption{Results for CHSH inequality measurement for maximally non-separable and separable modes. Theoretical results were obtained from entangled and product quantum sates. The experimental results were obtained from normalized intensity measurements of Figures \ref{nonsepmodes} and \ref{sepmodes}.}
\label{value_M}
\centering
\begin{tabular}{l|c|c|c|c|}
\cline{2-5}
                                                       & \multicolumn{2}{c|}{Maximally Non-Separable} & \multicolumn{2}{c|}{Separable} \\ \cline{2-5} 
                                                       & Theory             & Experiment            & Theory      & Experiment     \\ \hline
\multicolumn{1}{|l|}{$M(\alpha_1,\beta_1)$} & 0.707              & 0.679                   & 0.707       & 0.665            \\ \hline
\multicolumn{1}{|l|}{$M(\alpha_1,\beta_2)$} & 0.707              & 0.583                   & 0.000       & 0.000            \\ \hline
\multicolumn{1}{|l|}{$M(\alpha_2,\beta_1)$} & -0.707             & -0.679                  & -0.707      & -0.661           \\ \hline
\multicolumn{1}{|l|}{$M(\alpha_2,\beta_2)$} & 0.707              & 0.562                   & 0.000       & 0.000            \\ \hline
\multicolumn{1}{|c|}{$S$}                   & 2.828              & 2.503                   & 1.414       & 1.326            \\ \hline
\end{tabular}
\end{table}

For the separable mode the non violation of CHSH inequality were observed with $S=1.326$. This result are more close to the theoretically calculated once we have a better visibility in the MZIM.

\subsection{Kujala-Dzahafarov inequality for spin-orbit modes}

By using the normalized images intensities in Eq.\ref{d0} we obtained $\Delta_0 =0.143$. Therefore, the bound value of the Kujala--Dzahafarov inequalities (Eqs. \ref{eqcontex1} to \ref{eqcontex4}) is $S_{KD}<2(1+\Delta_0)=2.286$. Table \ref{skd} present the theoretical expected value of $S_{KD}$ for entangled state and the obtained results for spin--orbit modes calculated form Eqs. \ref{eqcontex1} to \ref{eqcontex4}. As can be seen, S$_{KD2} = 2.503 > 2.286$. This is a sufficient condition to infer the contextuality of spin--orbit modes. We notice a very good agreement between the quantum-theoretical prediction for the entangled state and the experimental results for nonseparable spin-orbit mode of a laser beam.   
% * <barros@sfsu.edu> 2018-07-10T00:45:02.994Z:
% 
% > By using the normalized images intensities in Eq.\ref{d0} we obtained $\Delta_0 =0.143$. Therefore, the bound value of the Kujala--Dzahafarov inequalitis (Eqs.\ref{eqcontex1} to \ref{eqcontex4}) is $S_{KD}<2(1+\Delta_0)=2.286$. Table \ref{skd} present the theoretical expected value of $S_{KD}$ for entangled state and the obtained results for spin--orbit modes calculated form Eqs. \ref{eqcontex1} to \ref{eqcontex4}. As can be seen, S$_{KD2} = 2.503 > 2.286$. This is a sufficient condition to infer the contextuality of spin--orbit modes. We notice a very good agreement between quantum theoretical prediction for entangled state and the experimental results for nonseparable spin--orbit mode of a laser beam. 
% 
% 
% See my comment above about redefining S_KD.
% 
% ^.

\begin{table}[!htp]
\centering
\caption{Results of theoretical expected value of $S_{KD}$ for entangled state (Theory) and the obtained results for spin--orbit modes (Experiment).}
\label{skd}
\begin{tabular}{c|c|c|}
\cline{2-3}
\multicolumn{1}{l|}{}                    & \textbf{Theory} & \textbf{Experiment} \\ \hline
\multicolumn{1}{|c|}{\textbf{$S_{KD1}$}} & 0.000           & 0.021                 \\ \hline
\multicolumn{1}{|c|}{\textbf{$S_{KD2}$}} & 2.828           & 2.503                 \\ \hline
\multicolumn{1}{|c|}{\textbf{$S_{KD3}$}} & 0.000           & 0.022                \\ \hline
\multicolumn{1}{|c|}{\textbf{$S_{KD4}$}} & 0.000           & 0.213                \\ \hline
\end{tabular}
\end{table}
% * <barros@sfsu.edu> 2018-07-10T00:47:40.270Z:
% 
% > \begin{table}[!htp]
% > \centering
% > \caption{Results of theoretical expected value of $S_{KD}$ for entangled state (Theory) and the obtained results for spin--orbit modes (Experiment).}
% > \label{skd}
% > \begin{tabular}{c|c|c|}
% > \cline{2-3}
% > \multicolumn{1}{l|}{}                    & \textbf{Theory} & \textbf{Experiment} \\ \hline
% > \multicolumn{1}{|c|}{\textbf{$S_{KD1}$}} & 0.000           & 0.021                 \\ \hline
% > \multicolumn{1}{|c|}{\textbf{$S_{KD2}$}} & 2.828           & 2.503                 \\ \hline
% > \multicolumn{1}{|c|}{\textbf{$S_{KD3}$}} & 0.000           & 0.022                \\ \hline
% > \multicolumn{1}{|c|}{\textbf{$S_{KD4}$}} & 0.000           & 0.213                \\ \hline
% > \end{tabular}
% > \end{table}
% 
% This table is a little strange in contrast with the previous table. The CHSH is not a single inequality, but a set of inequalities, the same as DK. The only difference between CHSH and DK is the lower boundary for the inequality because of the violation of the non-signaling condition.  This is why perhaps we should change the definition of S_KD, to make it clearer. 
% 
% ^.
 
The contextuality in this scenario reveals the conflict between nonseparable modes and noncontextuality theory in correlations between these degrees of freedom of a laser beam. This is analogue to the conflict between quantum mechanics and noncontextual realism. This result reinforces the idea that we can  use such systems to explore quantum properties; once the system presents an analogue mathematical structure to quantum systems it can be used to emulate single-photon experiments \cite{bellnos,kagawala, tripartitenos,environ,cqg2}. 
% * <barros@sfsu.edu> 2018-07-10T00:53:06.108Z:
% 
% > This result reinforces the idea that we can  use such systems to explore quantum properties; once the system presents an analogue mathematical structure to quantum systems it can be used to emulate single-photon experiments \cite{bellnos,kagawala, tripartitenos,environ,cqg2}. 
% 
% This sentence is a little weird, but I do not know how to re-write it properly. 
% 
% ^.

The revisitation on the experiment of violation of CHSH inequality for spin-orbit modes, beyond show an improvement in the violation, evidence,  in direct experimental measurements, the relation between Bell inequalities and noncontextuality discussed in Ref.\cite{cabello3}. 
% * <barros@sfsu.edu> 2018-07-10T00:54:14.424Z:
% 
% > The revisitation on the experiment of violation of CHSH inequality for spin-orbit modes, beyond show an improvement in the violation, evidence,  in direct experimental measurements, the relation between Bell inequalities and noncontextuality discussed in Ref.\cite{cabello3}. 
% 
% Again, it needs to be re-written. 
% 
% ^.

%%%%%%%%%CONCLUSIONS%%%%%%%%%%%%%%%%%%%%%%%%%%%%%%%
\section{\label{conclusions}Conclusions}

We have experimentally demonstrated the violation of the Kujala-Dzhafarov noncontextuality inequalities for spin-orbit mode of an intense laser beam. A nonseparable mode was prepared by a radial polarization converter, and we also showed an improvement of the violation of the CHSH inequalities by the nonseparable spin-orbit mode. Considering the relevance of contextuality in different scenarios, as universal quantum computation \cite{howard}, spin-orbit modes were shown to be an appropriate platform for the experimental investigation of contextuality. 
%%%%%%%%%%%%%%%%%%%%%%%%%%%%%%%%%%%%%
\begin{acknowledgments}
We would like to thanks Prof. Ernesto F. Galv\~ao by fruitful discussions. This research was supported by the Brazilian funding agencies Funda\c c\~ao Carlos Chagas Filho de Amparo \`a Pesquisa do Estado do Rio de Janeiro (FAPERJ), Conselho Nacional de Desenvolvimento Cient\'ifico e Tecnol\'ogico (CNPq), Coordena\c c\~ao de Aperfei\c coamento de Pessoal de N\'{\i}vel Superior (CAPES), and the National Institute of Science and Technology for Quantum Information.
\end{acknowledgments}

%\renewcommand{\bibname}{\leftline{Bibliografy}}
%\bibliographystyle{apsrev4-1}
%\bibliography{EnviromentInducedEntanglementExperiment}

%merlin.mbs apsrev4-1.bst 2010-07-25 4.21a (PWD, AO, DPC) hacked
%Control: key (0)
%Control: author (8) initials jnrlst
%Control: editor formatted (1) identically to author
%Control: production of article title (-1) disabled
%Control: page (0) single
%Control: year (1) truncated
%Control: production of eprint (0) enabled
\begin{thebibliography}{0}%
\makeatletter
\providecommand \@ifxundefined [1]{%
 \@ifx{#1\undefined}
}%
\providecommand \@ifnum [1]{%
 \ifnum #1\expandafter \@firstoftwo
 \else \expandafter \@secondoftwo
 \fi
}%
\providecommand \@ifx [1]{%
 \ifx #1\expandafter \@firstoftwo
 \else \expandafter \@secondoftwo
 \fi
}%
\providecommand \natexlab [1]{#1}%
\providecommand \enquote  [1]{``#1''}%
\providecommand \bibnamefont  [1]{#1}%
\providecommand \bibfnamefont [1]{#1}%
\providecommand \citenamefont [1]{#1}%
\providecommand \href@noop [0]{\@secondoftwo}%
\providecommand \href [0]{\begingroup \@sanitize@url \@href}%
\providecommand \@href[1]{\@@startlink{#1}\@@href}%
\providecommand \@@href[1]{\endgroup#1\@@endlink}%
\providecommand \@sanitize@url [0]{\catcode `\\12\catcode `\$12\catcode
  `\&12\catcode `\#12\catcode `\^12\catcode `\_12\catcode `\%12\relax}%
\providecommand \@@startlink[1]{}%
\providecommand \@@endlink[0]{}%
\providecommand \url  [0]{\begingroup\@sanitize@url \@url }%
\providecommand \@url [1]{\endgroup\@href {#1}{\urlprefix }}%
\providecommand \urlprefix  [0]{URL }%
\providecommand \Eprint [0]{\href }%
\providecommand \doibase [0]{http://dx.doi.org/}%
\providecommand \selectlanguage [0]{\@gobble}%
\providecommand \bibinfo  [0]{\@secondoftwo}%
\providecommand \bibfield  [0]{\@secondoftwo}%
\providecommand \translation [1]{[#1]}%
\providecommand \BibitemOpen [0]{}%
\providecommand \bibitemStop [0]{}%
\providecommand \bibitemNoStop [0]{.\EOS\space}%
\providecommand \EOS [0]{\spacefactor3000\relax}%
\providecommand \BibitemShut  [1]{\csname bibitem#1\endcsname}%
\let\auto@bib@innerbib\@empty
%</preamble>
\end{thebibliography}%


\begin{thebibliography}{1}

%The Problem of Hidden Variables in Quantum Mechanics.
\bibitem{origcontx} S. Kochen and E.P. Specker,  J. Math. Mech. 17, 59 (1967)

\bibitem{epr} A. Einstein, B. Podolsky, and N. Rosen, Phys. Rev., 47, 777–780 (1935).

%Simple Test for Hidden Variables in Spin-1 Systems.
\bibitem{kbcs} A. A.Klyachko, M. A. Can, S. Binicioglu, and A.S. Shumovsky, Phys. Rev. Lett.101, 020403 (2008).

%Quantum Contextuality in a Single-Neutron Optical Experiment. 
\bibitem{neutrons} Y. Hasegawa, R. Loidl, G. Badurek, M. Baron, and H. Rauch, Phys. Rev. Lett. 97 230401 (2006)

%State-independent experimental test of quantum contextuality.
\bibitem{iot}  Kirchmair, G. et al.  Nature 460, 494–497 (2009)

%T Experimental demonstration of quantum contextuality on an NMR qutrit. 
\bibitem{rmn} S.Dogra, K. Dorai, and K. Arvind, Phys. Lett. A 380,
1941–1946 (2006).

% Experimental Test of the Kochen-Specker Teorem with Single Photons.
\bibitem{photon1} Y.F.Huang,C.F Li, Y.S. Zhang, J.W. Pan, and G.C. Guo, 
Phys. Rev. Lett. 90, 250401 (2003).

% State-Independent Quantum Contextuality with Single Photons.
\bibitem{photon2} E.Amselem, M. Radmark, M. Bourennane, and A. Cabello, Phys.Rev. Lett. 103, 160405 (2009).

% Experimental Implementation of a Kochen-Specker Set of Quantum Tests. 
\bibitem{photon3} D’Ambrosio, V. et al.,Phys. Rev. X 3, 011012 (2013).

%Experimentally Testable State-Independent Quantum Contextuality
\bibitem{cabello1} A. Cabello, Phys.Rev. Lett.101, 210401 (2008).

%Spin-orbit hybrid entanglement of photons and quantum contextuality
\bibitem{padget1}  E. Karimi, J. Leach, S. Slussarenko, B. Piccirillo, L. Marrucci, L. Chen, W. She, S. Franke-Arnold, M. J. Padgett, and
E. Santamato, Phys. Rev. A 82, 022115 (2010)

%Contextuality and Indistinguishability
%\bibitem{acacio1} Jos\'e Acacio de Barros, Federico Holik, and D\'ecio Krause, Entropy  19, 435 (2017).

% Contextuality of identical particles
%\bibitem{kurzynski1} Pawel Kurzy\'{n}ski, Phys. Rev. A 95, 012133 (2017).

%
\bibitem{cabello2} Ad\'{a}n Cabello, Jos\'{e} Estebaranz, Guillermo Garc\'{\i}a-Alcaine, Phys. Lett. A, 212, 183-187 (1996).

%Computational Power of Correlations
\bibitem{compbcor} J. Anders and D. E. Browne, Phys. Rev. Lett. 102, 050502 (2009).

%Proposed Experiment to Test Local Hidden-Variable Theories
\bibitem{chsh} J. F. Clauser, M. A. Horne, A. Shimony, and R. A. Holt,
Phys. Rev. Lett. 23, 880 (1969).


%Wigner function negativity and contextuality in quantum computation on rebits, 
\bibitem{rebits} N. Delfosse, P. A. Guerin, J. Bian, and R. Raussendorf,Phys. Rev. X 5, 021003 (2015).


\bibitem{compmeasure} R. Raussendorf, Contextuality in measurement-based quantum computation, Phys. Rev. A 88, 022322 (2013).

%Reliable computation from contextual correlations
\bibitem{ernesto1} André L. Oestereich and Ernesto F. Galvão, 
Reliable computation from contextual correlations, Phys. Rev. A 96, 062305, (2017)


%Noncontextual wirings
\bibitem{wirings} B. Amaral, A. Cabello, M. Terra Cunha, L. Aolita, Phys. Rev. Lett. 120, 130403 (2018)
%DOI 10.1103/PhysRevLett.120.130403


%Contextuality in Three Types of Quantum-Mechanical Systems
\bibitem{kdl} Dzhafarov, E.N., Kujala, J.V. and Larsson, JA. Found Phys  45: 762. (2015)  
%https://doi.org/10.1007/s10701-015-9882-9

%Probabilistic Contextuality in EPR/Bohm-type Systems with Signaling Allowed.
\bibitem{kd2} Janne V. Kujala and Ehtibar N. Dzhafarov  Contextuality from Quantum Physics to Psychology: pp. 287-308 (2016). 
%https://doi.org/10.1142/9789814730617_0012 

%Linear optics and quantum maps
\bibitem{aiello1} A. Aiello, G. Puentes, and J. P. Woerdman
Phys. Rev. A 76, 032323 (2007)


%Topological phase for spin-orbit transformations on a laser beam
\bibitem{topo} C. E. R. Souza, J. A. O. Huguenin, P.Milman, and A. Z. Khoury, Phys. Rev. Lett. 99, 160401 (2007)


\bibitem{bellnos} C. V. S. Borges, M. Hor-Meyll, J. A. O. Huguenin, and A. Z.Khoury, Phys. Rev. A 82, 033833 (2010)

%Simulating Bell inequality violations with classical optics encoded qubits
\bibitem{Ledesma}M. A. Goldin, D. Francisco, and S. Ledesma, Journal of the Optical Society of America B 27, 779-786 (2010)


%Entanglement and classical polarization states
\bibitem{eberly1}X.-F. Qian and J. H. Eberly, Opt. Lett. 36, 004110 (2011).



%Classical hypercorrelation and wave-optics analogy of quantum superdense coding
\bibitem{kagawala} K. H. Kagalwala, G. Di Giuseppe, A. F. Abouraddy, and B. E.A. Saleh, Nat. Photon. 7, 72 (2012)


%Tripartite nonseparability in classical optics
\bibitem{tripartitenos} W. F. Balthazar, C. E. R. Souza, D. P. Caetano, E. F. Galv\~ao, J.A. O. Huguenin, and A. Z. Khoury, Opt. Lett. 41, 5797 (2016)


%
\bibitem{environ} M. H. M. Passos, W. F. Balthazar, A. Z. Khoury, M. Hor-Meyll, L. Davidovich, J. A. O. Huguenin, Phys. Rev. A \textbf{97}, 022321 (2018).


%Quantum key distribution without a shared reference frame
\bibitem{cript1} C. E. R. Souza, C. V. S. Borges, A. Z. Khoury, J. A. O. Huguenin, L. Aolita, and S. P. Walborn, Phys. Rev. A 77,
032345 (2008).

%Complete experimental toolbox for alignment-free quantum communication.
\bibitem{cript2} V. D’Ambrosio, E. Nagali, S. P. Walborn, L. Aolita, S. Slussarenko, L. Marrucci, and F. Sciarrino, Nat. Commun. 3, 961
(2012).

%Implementing the Deutsch algorithm with polarization and transverse spatial modes
\bibitem{lg1} A. N. de Oliveira, S. P. Walborn, and C. H. Monken, J. Opt. B:Quantum Semiclassical Opt. 7, 288 (2005)


%A Michelson controlled-not gate with a single-lens astigmatic mode converter
\bibitem{lg2} C. E. R. Souza and A. Z. Khoury, Opt. Express 18, 9207
(2010).

% . Using Polarization to Control the Phase of Spatial Modes for Application in Quantum Information. 
\bibitem{lg3}  W. F. Balthazar,  D. P. Caetano, ; C. E. R. Souza, J. A. O. Huguenin, Brazilian Journal of Physics, v. 44, 658-664 (2014).

%Vector vortex implementation of a quantum game
\bibitem{qg1} A. R. C. Pinheiro, C. E. R. Souza, D. P. Caetano, J. A. O.Huguenin, A. G. M. Schmidt, and A. Z. Khoury, J. Opt. Soc.
Am. B 30, 3210 (2013).

%  Simultaneous Quantum Duel.
\bibitem{qg2}  W. F. Baltazar, M.H. Passos, A.G.M. Schmidt, J. A. O.  Huguenin; Journal of the Physical Society of Japan, 84, 124002, (2015).

%Teleportation of a controllable orbital angular momentum generator
\bibitem{telep1} L. Chen and W. She, Phys. Rev. A 80, 063831 (2009)

%Remote preparation of single-photon "hybrid" entangled and vector-polarization States.
\bibitem{telep2} J. T. Barreiro, T. C. Wei, and P. G. Kwiat, Phys. Rev. Lett. 105,030407 (2010)

%Quantum teleportation in the spin-orbit variables of photon pairs
\bibitem{telep3} A. Z. Khoury and P. Milman, Phys. Rev. A 83, 060301
(2011)

%Shifting the quantum-classical boundary: theory and experiment for statistically classical optical fields
\bibitem{eberly2} Xiao-Feng Qian, Bethany Little, John C. Howell, and J. H. Eberly, Optica 2, 611-615 (2015) 


%Classical entanglement in polarization metrology
\bibitem{aiello2} Falk Töppel, Andrea Aiello1, Christoph Marquardt, 
Elisabeth Giacobino, and Gerd Leuchs,New Journal of Physics 16, 073019 (2014) 
%10.1088/1367-2630/16/7/073019




%Experimental contextuality in classical light
\bibitem{Ckbcs} Tao Li, Qiang Zeng, Xinbing Song, and Xiangdong ZhangSci.  Rep. 7, 44467 (2017).

%Contextuality-by-Default 2.0: Systems with Binary Random Variables.
\bibitem{kd1} Dzhafarov E.N., Kujala J.V. (2017)  In: de Barros J., Coecke B., Pothos E. (eds) Quantum Interaction. QI 2016. Lecture Notes in Computer Science, vol 10106. Springer, Cham

% Aspect's doctoral dissertation
\bibitem{aspectdr} A. Aspect. Trois tests expérimentaux des inégalités de Bell par corrélation de polarisation de photons. Physique Atomique [physics.atom-ph]. Université Paris Sud - Paris XI, 1983. Français. <tel-00011844>

%Long-distance Bell-type tests using energy-time entangled photons
\bibitem{loop1} E. Santos, Phys. Lett.A 212, 10 ~1996.

%
\bibitem{loop2}W. Tittel, J. Brendel, N. Gisin, and H. ZbindenPhys. Rev. A 59, 4150

%Transverse-mode beam splitter of a light beam and its application to quantum cryptography
\bibitem{mzim} H. Sasada and M. Okamoto, Phys. Rev. A 68, 012323 (2003).

%Quantum and classical separability of spin--orbit laser modes
\bibitem{Kaled} Pereira, L. J., Khoury, A. Z., Dechoum, K.,  Phys. Rev. A, 90, 053842, 2014. 

\bibitem{environ} M. H. M. Passos, W. F. Balthazar, A. Z. Khoury, M. Hor-Meyll, L. Davidovich, and J. A. O. Huguenin, Phys. Rev. A 97, 022321 (2018). 


\bibitem{cqg2} Balthazar, W. F., Caetano, D. P,  Souza, C. E. R., Huguenin, J. A. O, Brazilian Journal of Physics, 44, 658-664 (2014).

\bibitem{cabello3}Kurzyński, P., Cabello, A., and Kaszlikowski, D., Phys. Rev.Lett. 112, 100401 (2014).

\bibitem{howard}Howard, M., Wallman, J., Veitch, V. \& Emerson, J, Nature, 510, 351–355(2014).


\bibitem{pawel}Kaszlikowski, D., Kurzy\'nski, P. (2016). In: de Barros J., Coecke B., Pothos E. (eds) Quantum Interaction. QI 2016. Lecture Notes in Computer Science, vol 10106. Springer, Cham





\end{thebibliography}

\end{document}